\documentclass[12pt]{iopart}
\usepackage[font=small]{caption} 
\usepackage{bm}
\usepackage[printwatermark]{xwatermark}
\usepackage{xcolor}
\usepackage{graphicx}             
\usepackage{txfonts}
\usepackage{fullpage}             
\usepackage{url}
\usepackage{placeins}
\usepackage{dcolumn}
\usepackage{bm}
\usepackage[printwatermark]{xwatermark}
\usepackage{xcolor}
\usepackage{lipsum}
\usepackage{datetime}
\usepackage{url}
\usepackage{citesort}
\usepackage[numbers,sort&compress]{natbib}

\DeclareMathAlphabet{\pazocal}{OMS}{zplm}{m}{n}
\begin{document}
\title{Local Field Potential Journey into the Basal Ganglia}
\author{Eitan E. Asher$^1$, Maya Slovik$^{2,3}$,
Rea Mitelman$^{2,3}$, Hagai Bergman$^{2,3}$, Shlomo Havlin$^1$,  Shay Moshel$^{1,2,3,4}$}

\address	{
$^1$Department of Physics, Bar-Ilan University, Ramat Gan, Israel\\
$^2$Department of Medical Neurobiology, The Hebrew University-Hadassah Medical School, Jerusalem, Israel\\
$^3$ The Edmond and Lily Safra Center for Brain Sciences, The Hebrew University, Jerusalem, Israel
$^4$ Nuclear Research Center Negev, Beer-Sheva, Israel
}

\ead{shaymoshel@gmail.com}

\begin{abstract}
	
Local field potentials (LFP) in the basal ganglia (BG) have attracted considerable research and clinical interest. Over the years, much debate has been occurred regarding the origin of this signal. Whether or not it is a local sub-threshold phenomenon, reflecting, for example, the synaptic input to neurons in the recording area, or a flow of electrical signals generated by simultaneously firing neurons in the cerebral cortex that obeys the Maxwell equations (volume conduction). Here, we recorded simultaneously in the non-human primate (NHP) brain LFP’s from two loci in the cerebral cortex: the dorsolateral prefrontal (DLPFC) and the primary motor (M1) cortex, as well as multiple sites in the BG nuclei: the striatum, globus pallidus, and subthalamic nucleus. All recordings were taken while the NHPs were awake but not engaged in any behavioral tasks. By developing and applying novel methods to identify significant cross correlations (potential links) while removing ”spurious” correlations, we developed a tool that may help in discriminating between synaptic inputs (information flow) and volume conduction. Our analysis reveals that there are two main paths of BG field potentials, propagating with two different time lags from the primary motor cortex and the DLPFC. These two anatomical pathways are found to be differently affected by the two mechanisms of volume conduction and information flow. 

\end{abstract}

\maketitle 

\section{Introduction}
Broad band field potential becomes an important and powerful tool to understand many neuronal functions, abnormalities and clinical phenomena.
Behavioral studies use local field potential (LFP) activity to describe cognitive occurrences \cite{buzsaki2012origin,wimmer2016transitions,widge2019prefrontal}
and sensory response \cite{henrie2005lfp,bosman2012attentional, carmichael2017gamma}.
Brain machine interface tools have been developed based on LFPs \cite{jackson2016decoding} and electroencephalogram (EEG) signals \cite{muller2008machine}.
Subthalamic nucleus LFPs has been found to be coupled with EEGs related to cortical somatosensory areas \cite{marsden2001subthalamic}, and neural discharge rates found to be coupled with LFPs signals in the cerebral cortex during voluntary movements \cite{donoghue1998neural}.
Brain diseases such as Parkinson disease (PD) and schizophrenia, have been found to be correlated with the increase of brain abnormal oscillations in neuronal signals, as single-unit, multi-unit activity and LFPs in the Basal Ganglia (BG) and cerebral cortex \cite{rodriguez2001subthalamic,moshel2013subthalamic, slovik2017ketamine}. 

 Classical LFP  frequency domain is [0.1-70] Hz, where frequencies higher than 100 Hz are described as high frequency oscillations in LFP  \cite{ozturk2021electroceutically}. 
However, Yuval-Greenberg et. al. \cite{yuval2008transient} shows possible confounding factors in the high frequency regime of LFP, which are most probably due to spikes that have their maximal power around 500-1500 Hz. Although LFP oscillations have been thought to imply spike synchronization \cite{brown2005basal, hammond2007pathological,ozkurt2011high, moshel2013subthalamic}, they are more likely to represent sub-threshold phenomena as synaptic inputs \cite{belitski2010sensory,buzsaki2012origin} which is probably correlated with spike activity.
Broadband field potentials have been also found to be correlated with spontaneous cerebral low-frequency blood oxygen level-dependent (BOLD) fluctuations 
\cite{pan2011broadband, hermes2017neuronal}. 
Hermes et al., \cite{hermes2017neuronal} described that, the two measures, the BOLD amplitude and electrocortiogram (ECoG) broadband power are correlated in cerebral cortex (V1, V2 and V3). More specifically, the BOLD amplitude and alpha power (8–13 Hz) are negatively correlated, and the BOLD amplitude and narrow--band gamma power (30–80 Hz) are uncorrelated. The authors pointed out that the two measures, provide complementary information about human brain activity, and they infer those features of the field potential that are uncorrelated with BOLD arise largely from changes in neuronal synchrony, rather than, level of direct neuronal activity.

Although broadband field potential plays a key role for analyzing brain behavior and it is easy to record, the origin of LFP remains under debate. Many studies suggest that this signal carries valuable neuronal information as the local synaptic inputs \cite{maier2011infragranular, lempka2013theoretical,okun2010subthreshold,brown2005basal, linden2010intrinsic}.
On the other hand, there are studies that describe these signals as propagative electromagnetic waves which behave as a volume conductance, obeying the Maxwell equations \cite{tenke2012generator,hermes2017neuronal,parabucki2017volume,einevoll2013modelling}. 
For example, Parabucki et. al. \cite{parabucki2017volume} described strong
LFP signals coupling between the cerebral cortex, during whisker stimulation in a mice, with a dissected part of the olfactory bulb. They recorded simultaneously the cerebral cortex and the dissected part and describe that the LFP “passed” from the cortex through the dissected part as a volume conducted signal. An intermediate study suggests that these signals have mixed characters of both volume conduction and a summation of the extracellular fields \cite{buzsaki2012origin}. 
L.Lalla et al \cite{lalla2017local} studied LFP oscillations and spike activity simultaneously recorded in the dorsolateral striatum (DLS) and the forelimb primary somatosensory cortex (S1) of rats during the execution of a stereotyped running sequence. They observed an increase in rhythmic activity of LFP in the theta band during the run compared to the rest period. They claimed that a common external source should be mostly accounted for the interdependence of LFPs signals recorded in the cortex and the striatum. One of their prominent results was that these theta oscillations observed in striatal LFPs were largely “contaminated” by volume-conducted signals. 
 However, despite of these remarkable studies, to date we still do not have a clear picture for describing the activity of broadband field potentials, whether it is a volume conducted electromagnetic waves that spreads inside the deep layers in the brain or they carry a valuable information as synaptic inputs. 
 Moreover, if it is an intermediate phenomenon, it is of much interest, what would be the rate of volume conduction and synaptic inputs, from LFP signal, whether there are pure patterns of volume conduction or synaptic input realizations, whether they are long, continuous, or having short time duration?   
To explore this issue, we map LFP signals from the cerebral cortex trough the basal ganglia nuclei, where we recorded simultaneously the signals from the cerebral cortex and signals in the deep layers of the basal ganglia (BG) nuclei, of non-human primates (NHP) during wake. Using micro-electrodes, we recorded the full electrophysiological signal (raw-data) in the range of [1, 6000] Hz, with 1-7 electrodes which were fixed in the deep cortical layers (pyramidal cells) for all recording session, see Fig.\ref{fig:brainElects}(A).
An eighth electrode has been moved towards to the BG nuclei. The neural activity of all the electrodes has been simultaneously recorded every 200 $\mu$ m, for a duration of 120 seconds, since the entrance of the 8th electrode to the cerebral cortex (frontal lobe) and while passing through: the striatum, globus pallidus (external - GPe, internal - GPi), internal capsule and the subthalamic nucleus. 
 To get LFP signal we filtered the full signal (raw data) to 1-80Hz. The simultaneous LFP recordings were analyzed to reveal the relationship between neuronal activity in the cerebral cortex and the deep BG brain layers. 

\section{Methods}
\subsection{Animals, surgery and MRI}
We conducted the experiments on one African Green monkey (Cercopithecus aethiops, female, weighing 3.3 Kg). All procedures were conducted in accordance with the Hebrew University guidelines for animal care and the National Institute of Health Guide for the Care and Use of Laboratory Animals. The Hebrew University is an AAALAC (Association for Assessment and Accreditation of Laboratory Animal Care) approved institute. 
The monkey was trained to sit quietly in a primate chair. After training, the animal underwent a surgical implantation of a Cylux MRI compatible 27*27 mm recording chamber (Alpha-Omega, Israel) and head holder (Crist Instruments, MD). The location of the chamber was determined by a primate stereotaxic atlas \cite{contreras1981stereotaxic, miocinovic2007stereotactic} to be above the right primary motor cortex (M1), globus pallidus (Gp) and the subthalamic nucleus (Stn). A few days after surgery, an MRI scan was performed to determine its exact stereotaxic location.
\subsection{Recording and data acquisition}
During recording sessions, the animal was seated in a primate chair with its head and hands restrained but was free to move their trunk and legs. All recording were done during quite alert state. The verification of recording location was carried out separately for the target structures. M1 was verified by limbs motor response to 10-40 $\mu A$ stimulation. DLPFC was detected by its stereotactic and chamber.
The BG nuclei  complex: striatum, globus pallidus and the subthalamic nucleus were discriminated by their physiological features. In addition, the subthalamic nucleus, was verified by cellular response to passive limbs motor stimulation. Extra cellular data domain was amplified with a gain of 5K, band- pass filtered with a 1-6000 Hz four-pole hardware Butterworth filter and continuously sampled at 25 kHz.

\begin{figure}[ht!]
	\includegraphics[width=16.5cm, height=8.5cm ]{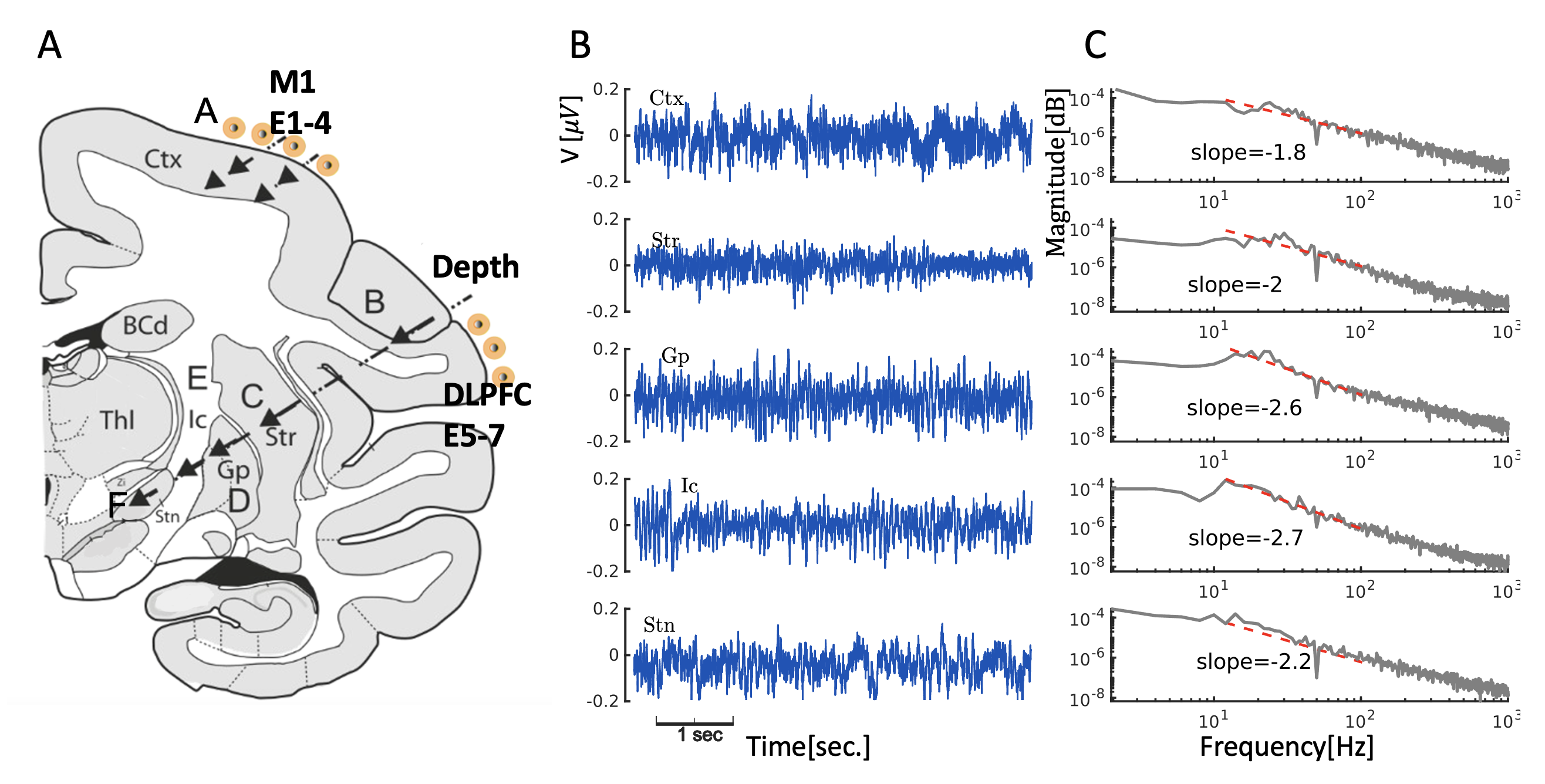}
	\caption{{\bf Anatomical and electrodes setup, raw data and power spectrum densities.}
	\textbf{A,} 
	Hemi-coronal illustration from Macaque atlas, at the level of anterior commissure (AC) - 6 mm. We placed 7 electrodes in the deep layers of the cerebral cortex. Electrodes 1-4 (M1) were permanently located on the right–hemisphere, primary M1- motor-cortex, and electrodes 5-7 (DLPFC) on the right hemisphere – dorso-lateral frontal cortex. One electrode aimed to the basal ganglia, passing through the recording sites in: the cerebral cortex (Ctx), striatum (Str), globus pallidus (Gp) internal capsule (Ic) and subthalamic nucleus (Stn). Each site has been recorded for time duration of 120 seconds. Labels in cartoon: A to F mark the recorded area. \textbf{B,}  Examples of 5 seconds records of raw local field potential data from shallow to deep loci: Ctx, Str, Gp, Ic and Stn, and \textbf{C,} corresponding power spectral density, after applying a notch filter to remove the 50Hz band.
	}	
	\label{fig:brainElects}
\end{figure}
\FloatBarrier

 The data has been stored by a data acquisition system (Alpha-Map, Alpha-Omega, Israel) for offline analysis. The behavior has been monitored by a computerized video surveillance system (GV-650, GeoVision, Taiwan;) connected to four infrared digital cameras. 
We used two arrays of four glass-coated tungsten microelectrodes (impedance measured at 1 kHz, range 0.2-0.6 MOhm), confined within a cylindrical metal guide (1.36 mm internal diameter). The electrodes were positioned (M1 electrodes) in the area of the primary motor cortex and the other E5-7 electrodes were fixed at frontal cortex and E8 aimed to the basal ganglia nuclei , see Fig.\ref{fig:brainElects}(a). This was done using the double Microdriving Terminal and Electrode Positioning System (Double MT and EPS, Alpha-Omega, Israel), that allows independent movement of each electrode. 
During the recording procedure, we fixed E1-7 electrodes at the deep layers of the cerebral cortex (M1 and the frontal cortex above the BG) and a moving 8th electrode, has been advanced towards the BG nuclei (Figure 1A). We started simultaneous recordings of the Ctx electrodes (E1-7) when the moving electrode entered the cerebral cortex. We recorded at each location for time duration of two minutes, after waiting several seconds for stabilization. When we completed the specific site recording; we moved further the traveling electrode by 200 $\mu$m to the next deeper location.
In our data analysis we utilized Python 3.6 and custom-made MATLAB19a (R2019a) routines. Our database contains 13 trajectories from the basal ganglia nuclei.
	Data records are summed in Table \ref{Tab:1}.
      
\begin{table}
	\begin{center}
		\begin{tabular}{|c|c|c|c|c|c|}
			\hline
 Location & Ctx & Str &Gp &  Ic & Stn \\
 \hline
 Sites  & 210  &219 & 91 & 20 & 18  \\
 Pairs  & 1318 & 1331 & 561 & 140 & 126 \\
			\hline
		\end{tabular} 
		\caption{{\bf Database of the study.} 
			Number of recorded sites of each structure, and number of total cortex electrode-deep electrode pairs. The deep electrode (E8) was moved in 200$\mu$m steps from the cortex towards the BG. Each location that has been visited is a site. At each site, the signal that was recorded from the deep electrode has been analyzed using cross correlation against its matching time record of the E1-7 electrodes, resulting in the number of total pairs. Ctx - cortex, Str - striatum, Gp - globus pallidus, Ic - internal capsule, and stn - subthalamic nucleus.  }

	\label{Tab:1} 
    \end{center}
\end{table}
\FloatBarrier

\subsection{Cross-correlation and surrogate analysis to	identify significant interactions}

\begin{figure}[h!]
	\includegraphics[height=9cm, width=16.8cm]{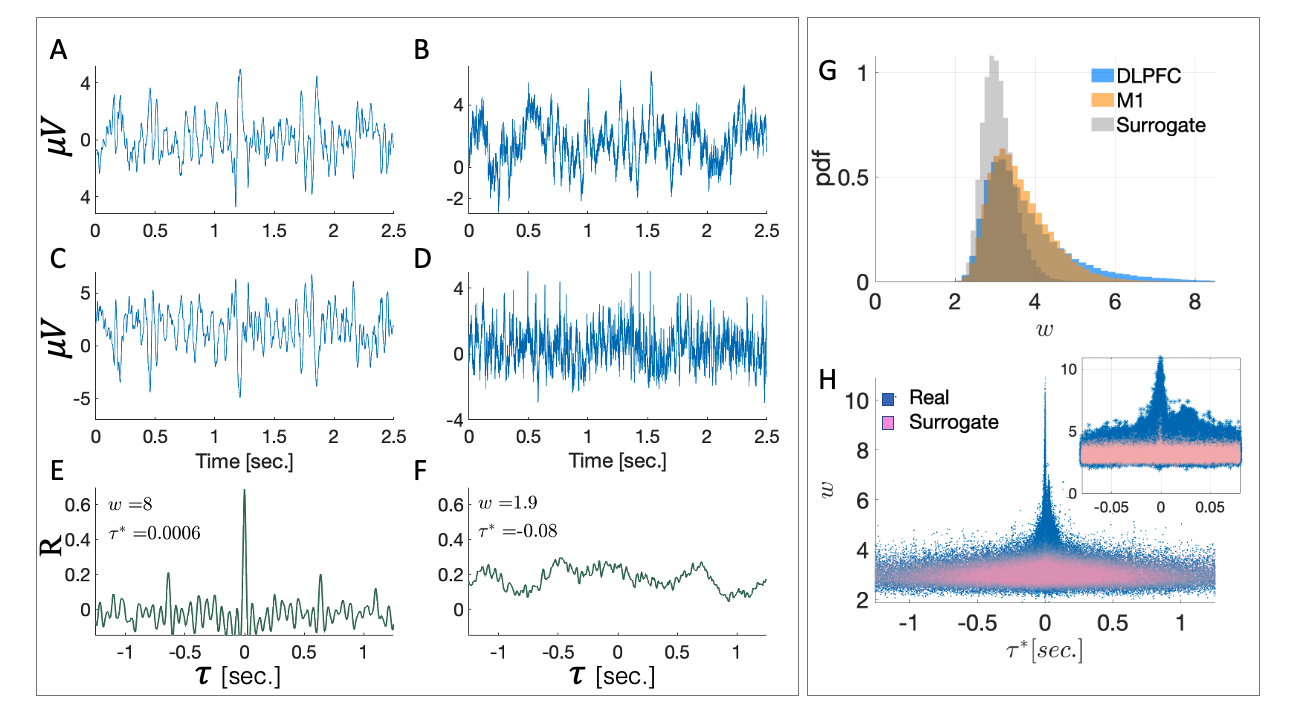}
	\centering
	\caption{{\bf Identification of significant interactions by correlations and surrogate analysis.}
		\textbf{A-F,} 
		 Demonstration of two pairs of signals (green curves in A-C and B-D) obtained simultaneously from different recording loci after applying a [1-80Hz] low-pass-filter to the raw LFP data. \textbf{E-F},  The cross correlation index $R$ as a function of the shift $\tau$ between the signals A vs. C and B vs. D, respectively. The signals from A and C yield a marked maximum $R$ at shift $\tau^\ast=\left.\tau\right\vert_{R(\tau)\equiv R_{\rm{max}}}^{}=0$, and $R(\tau)$ decays rapidly for $|\tau|>0$. \textbf{B} and \textbf{D} demonstrate uncorrelated signals, since, $R(\tau)$ in F shows fluctuating behavior without clear decay, i.e, uncorrelated signals. A significance value $w$ characterizes $R(\tau)$ by normalizing $R_{\rm{max}}$ by the mean and standard deviation of $R(\tau)$ (Eq.~\ref{w_eq}). Correspondingly, we obtain a higher $w$ value for panel E ($w=8$) than for panel F ($w=1.9$). \textbf{G,} distribution of $w$ for all data, taken from electrodes and DLPFC, compared to surrogate signals. For the real data, we show $w$ for all time windows that were analyzed. Surrogate signals are signals that were randomly chosen from different electrodes, different nuclei, and also at different times.  Surrogate signals yield $w$ values distribution that is significantly lower than the distribution of real signals; maximum $w$ value for surrogate signals is around 4. Panel \textbf{H,} shows that highest $w$ values are observed for $\tau^\ast\approx 0$. In this scatter plot we show 5000 samples of $w$ vs. $\tau^\ast$ for real data (blue dots) and surrogate data (red dots). Real signals are taken at the same time, comparing deep and Ctx signals, whereas surrogate pairs were chosen randomly at different times. Clearly, higher $w$ values are obtained for real signals for $\tau^\ast\approx 0$. In contrast, surrogate analysis does not lead to higher $w$ values around $\tau^\ast\approx 0$ and shows a uniform $w$ vs. $\tau^\ast$ distribution. The inset shows a zoomed-in view of $\tau^\ast \in [-0.08-0.08]$ region.
	}
	\label{fig: Methods}
\end{figure}

We used cross-correlation analysis to quantify the similarities between two signals, X and Y, namely from recorded cerebral cortex electrodes (E1-7) and BG sites (E-8), see Fig. \ref{fig: corrs}.
 We divide both signals ${x}$ and ${y}$ into $N_L$-overlapping segments $\nu$ of equal length L=2.5 seconds. We choose an overlap of L/4=0.625s.
 Thus, for each 120s record we get $N_L=63$ time windows.
 To remove constant trends in the data, the signal in each segment $\nu$ is subtracted by its mean. This normalization assures that the coupling between the signals ${x}$ and ${y}$ is affected mainly by their variations from the average. Next, we calculate the cross correlation function, $R^\nu_{xy}(\tau) = \sum_{\tau = -n/2}^{n/2} x^\nu(n+\tau)* y^\nu(n)$ within each segment $\nu=1,...,N_L$.
To distinguish between significant and non-significant interactions between signals ${x}$ and ${y}$, we study how their cross-correlation $R$ decays when shifting the signals against each other. The more the cross-correlation peak is above or below the background level, the more significant is the link or the mutual coupling.
 Panels (A-F) in Fig. (\ref{fig: Methods}) show examples (A,C and B,D ) of R versus time shift $\tau$ (E and F) and suggest that more synchronized (A,C) signals (with higher R values) have a marked decay of R($\tau$) (panel E) that are not seen (panel F) for non-synchronized signals (B and D) . To better quantify this observation, we define
 a Z-score-like index ‘w’ to determine the correlation strength or link strength. The index $w$ of maximal $R$ value is defined as the number of standard deviations above (or below) the mean over the shifting range. In our case, the parameter $w$ normalizes the maximum value, peak value - $R_{max}$ (detected at a particular time shift $\tau^*$) of the shifted-cross-correlation and mean, by the  standard deviation of R($\tau$). Informally, $w$ gives an estimate of how much $R_{max}$ “stands out” from the background, see Fig. \ref{fig: Methods}(E). Specifically, we ask how many standard deviations the value of the peak is farther from the background noise. That is,
 \begin{equation}
w ^{i,j}=  \frac{(|R_{max}|)- \langle R \rangle)}{\sigma(R)} 
\label{w_eq}
 \end{equation}

where $\langle R(\tau)\rangle$ ($\sigma(R(\tau) )$) are the mean and standard deviation of $R(\tau)$ respectively.
From Eq. (\ref{w_eq}) one can see that if $R_{max}$ is much larger than the background, $w$ is higher, indicating a more significant coupling between signals $x$ and $y$. On the other hand, if $R_{max}$ does not stand out from the $R(\tau)$ background, $w$ is lower and implies non-significant $x – y$ interaction. Another way to test significant links is by comparing cross correlations of simultaneous signals from different locations to those taken at different times which are expected to be  non synchronized. Figure (\ref{fig: Methods})G) shows results obtained from 10,000 pairs of signals, where each pair of signals is either taken at the same time (simultaneously, real” data) or at different times (“surrogate” data).
It can be seen from Fig. \ref{fig: Methods}(G), that for surrogate data one obtains low values of w (just above 4), while for real data w can reach values of 8. Moreover, as seen in Fig. \ref{fig: Methods} (H), high $w$ values for the real data are usually detected for small shifts $\tau^*$ (of the order of few tens of milliseconds), which is consistent with the fact that brain waves generally propagate rather fast \cite{zhang2018theta,sato2012traveling}.
Combining the two conditions, we consider $x – y$ interactions as significant only if $\tau^\ast\in [-0.05,0.05]$ seconds and $w>4.5$. Table \ref{Tab:2} summarizes the real links fraction at each loci.
\FloatBarrier

\FloatBarrier

\begin{table}
	\begin{center}
		\begin{tabular}{|c|c|c|c|c|c|}
			\hline
			Location & Ctx & Str & Gp & Ic & Stn \\
			\hline
			M1            & 0.08  &0.14 & 0.14 & 0.13 & 0.08  \\
			DLPFC    & 0.51 & 0.24 & 0.09 & 0.24 & 0.04 \\
			\hline
		\end{tabular} 
		\caption{{\bf Fraction of real links per nuclei and location.} 
			Fraction of real links calculated as the ratio between the number of real links divided by the total number of recorded time windows in each site. The real links were chosen by the conditions, $\tau^\ast \in [-0.05; 0.05]$ seconds and $w>4.5$. }
		\label{Tab:2} 
	\end{center}
\end{table}
\FloatBarrier
\section{Results}
\label{sec: results}

\begin{figure}[h!]
	\includegraphics[height=16.8cm ,width=16.5cm]{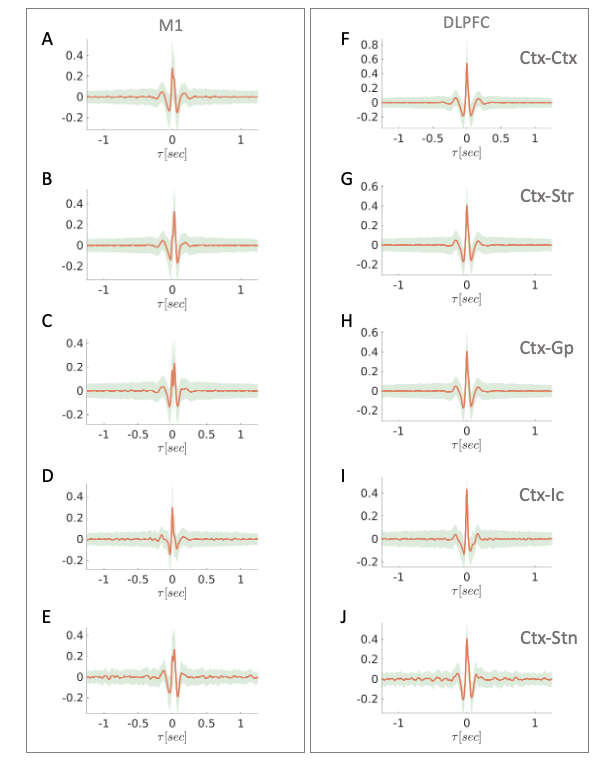}
	\caption{ \textbf{ Cross-correlations.} The cross-correlations of the M1 and DLPFC electrodes with the deep electrode located at the DLPFC and  in the BG nuclei. Left column (A, C, E, G, I) shows the cross-correlations between DLPFC-Ctx, DLPFC-Str, DLPFC-Gp, DLPFC-Ic and DLPFC-Stn respectively. On the right column, we show the same interactions for the DLPFC - Ctx.
	  Here, we show only correlations with $w > 4.5$ and $\tau^\ast<0.05$ sec., of the time windows that were analyzed. In red is the mean, and the shaded area represents the standard deviation. } 
	\label{fig: corrs}
\end{figure}

{\bf Correlations, scores and $\tau^*$ distributions.}
Only part of realizations were found to be synchronized and found to be significant links, see Table 2: in DLPFC-BG pairs the percentage of "successful links" has been found to be: between 4\%-51\%; for M1-BG pairs the significant links are between 8\%-51\% of the total time windows (see Table \ref{Tab:2}). Figure  \ref{fig: corrs}  describes the strong links with significant cross-correlations.
We define the time lag $\tau^\ast$ at which $R(\tau)$ is maximal, as the time delay between the signal pairs at $x$ and $y$. Whether $\tau^\ast$ is positive or negative, determines the location in which the signal was detected first and points to the direction of interaction. In the case of positive $\tau^\ast$, the interaction starts at the cortex and moves to the basal ganglia, whereas in the case of negative $\tau^\ast$, the interaction follows the opposite direction.

\begin{figure}[h]
	\includegraphics[width=16.5cm]{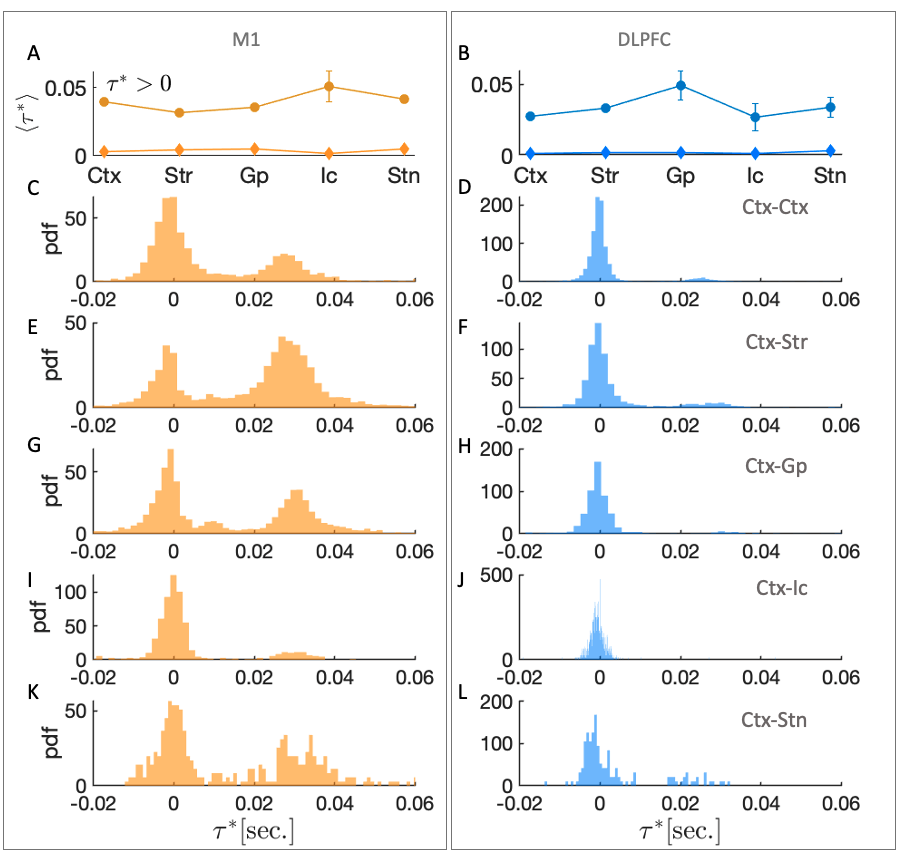}
	\centering
	\caption{{\bf Basal ganglia - cerebral cortex, average time delay and $\tau^\ast$ distribution. } \textbf{A-B,} Average $\tau^\ast$'s of nuclei, of interactions with $\tau^\ast>0$. We show the mean time delays of the two signals of the electrodes sets, i.e., between each (A) M1, (B) DLPFC, and the moving electrode. $\tau^\ast>0$ corresponds to the case in which the signal initiated (or earlier) at the cortex. This consists of $80\%$ of the real links. 
	\textbf{C-L} The distribution of $\tau^\ast$ for all cerebral cortex-BG nuclei interactions. Note that the distributions are bi-modal, suggesting two types of links. 
 }  
	\label{fig:tauresults}
\end{figure}

 We find that when looking at real links in which $w > 4.5$, the average $\tau^\ast$ is almost the same for the M1 and DLPFC as seen in Figure \ref{fig:tauresults}, A-B.  Furthermore, both cortical area distributions display a bi-modal distribution, Fig. \ref{fig:tauresults}, C-L.
While the BG-DLPFC (Fig. \ref{fig:tauresults} left panels) interactions are mostly dominated by the first mode ($\tau^\ast \sim 0$),
the BG-M1 interactions (Fig. \ref{fig:tauresults} right panels) exhibits both first and second modes time delays. 

\begin{figure}[h]
	\centering
	\includegraphics[width=16.5cm, height=15cm]{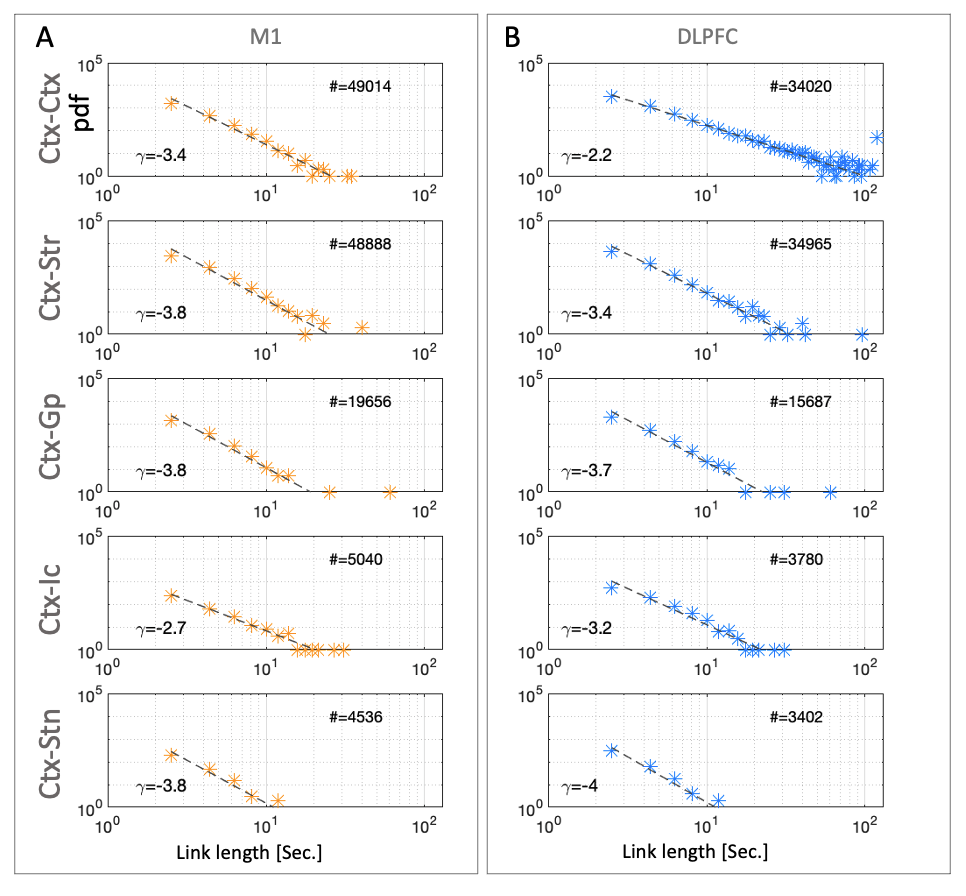}
	\caption{
		{\bf Distribution of links duration of different loci. } 
		Left and right columns show PDFs of links time duration of M1 and DLPFC electrode sets respectively. Each site record (120 sec.), is divided in to 2.5 seconds overlapping windows, with $25\%$ overlap (0.625 seconds). Then, we analyze the cross-correlation within each time window segment between two signals, one in the deep electrode and the second its matching time record of the electrode in the cortex, and $w$ is calculated. If $w>4.5$, we assume that the link is real.
	 We then find the duration of the consecutive real links that were formed with the cortex electrodes at all sites of the record. The results presented in log-log scale, suggest a power law distribution behavior (exponent $\gamma$ is given). It can be seen that the declination of real links duration distribution is higher as the nuclei is deeper. Note also, that for most nuclei links duration are longer for the DLPFC frontal lobe electrode set compared to the M1 set. This suggests that spatial distance affects the duration of the LFP signals.	 
	  Exponents show consistent increase with depth, except Ic, which consists of white matter in which signals might propagate more rapidly. }
	\label{fig:linklengths}
\end{figure}

{\bf Link time distribution.} Another question of interest is what is the time duration or "lifetime" of a significant link? 
As before, we divide each recording time period at each site (120 s) into 2.5 s overlapping windows, with a fraction of 0.25 s window length overlap. Then, by calculating the cross-correlation of the signal of a given site with its simultaneous time signal in the cortex electrode, we find the link strength $w$ by applying Eq. (\ref{w_eq}). If for a time window of 2.5 seconds, $w>4.5$, we regard the two signals as linked in this window. 
Then, we find the time duration of this link by counting the number of successive windows that are correlated. Next, we plot in Fig. \ref{fig:linklengths} their lifetime distribution for all the nuclei interactions that were probed. We find that, as the probe moves towards deeper nucleus, there is a clear decrease in the link length. Upper area links maximal duration as M1-DLPFC and DLPFC-DLPFC were 20 and 40 seconds respectively. M1-Str and DLPFC-Str are 25 sec and 30 seconds respectively. M1-GP and DLPFC-GP are both around 20 sec, either the Ic links. Both M1/DLPFC - Stn links were 10 seconds time duration. 
 
{\bf Modes analysis.} In Fig. \ref{fig:tauresults}, we observe a bi-modal distribution of $\tau^\ast$ which is significantly more pronounced in the M1-GP links, supporting the hypothesis of the existence of two transfer mechanisms. Both paths have similar $\tau^\ast$'s but the the DLPFC-GP links are dominated by $\tau^\ast$ centered around zero time lag (Fig. \ref{fig:tauresults} right panels) while the M1-GP paths is influenced almost equally also by a larger $\tau^\ast$ typically of 30 milliseconds (Fig. \ref{fig:tauresults} left panels). The links with $\tau^\ast \sim 0$ suggest, a volume conduction mode, and the second mode, with larger $\tau^\ast$ could indicate information transfer through synaptic inputs. As shown above, every significant link is composed of several consecutive sub-links of 2.5 seconds each.

 To support further the above hypothesis of the existence of two modes we will test next if all or at least most of sub-links of a specific link belong to the same mode or if we have a random mixture of modes in the same link. This test is a crucial since if the links are mainly unique, i.e., one of the two mechanisms, it is expected that almost all (almost -- because there could be some outliers due to statistical errors) sub-links in a specific link should have the same $\tau^\ast$, as indeed shown below, as seen in Fig. \ref{fig:modes_composition}. This finding could highly support our hypothesis of the existence of two transfer mechanisms.
 
\begin{figure}[h]
	\centering
	\includegraphics[scale=0.39]{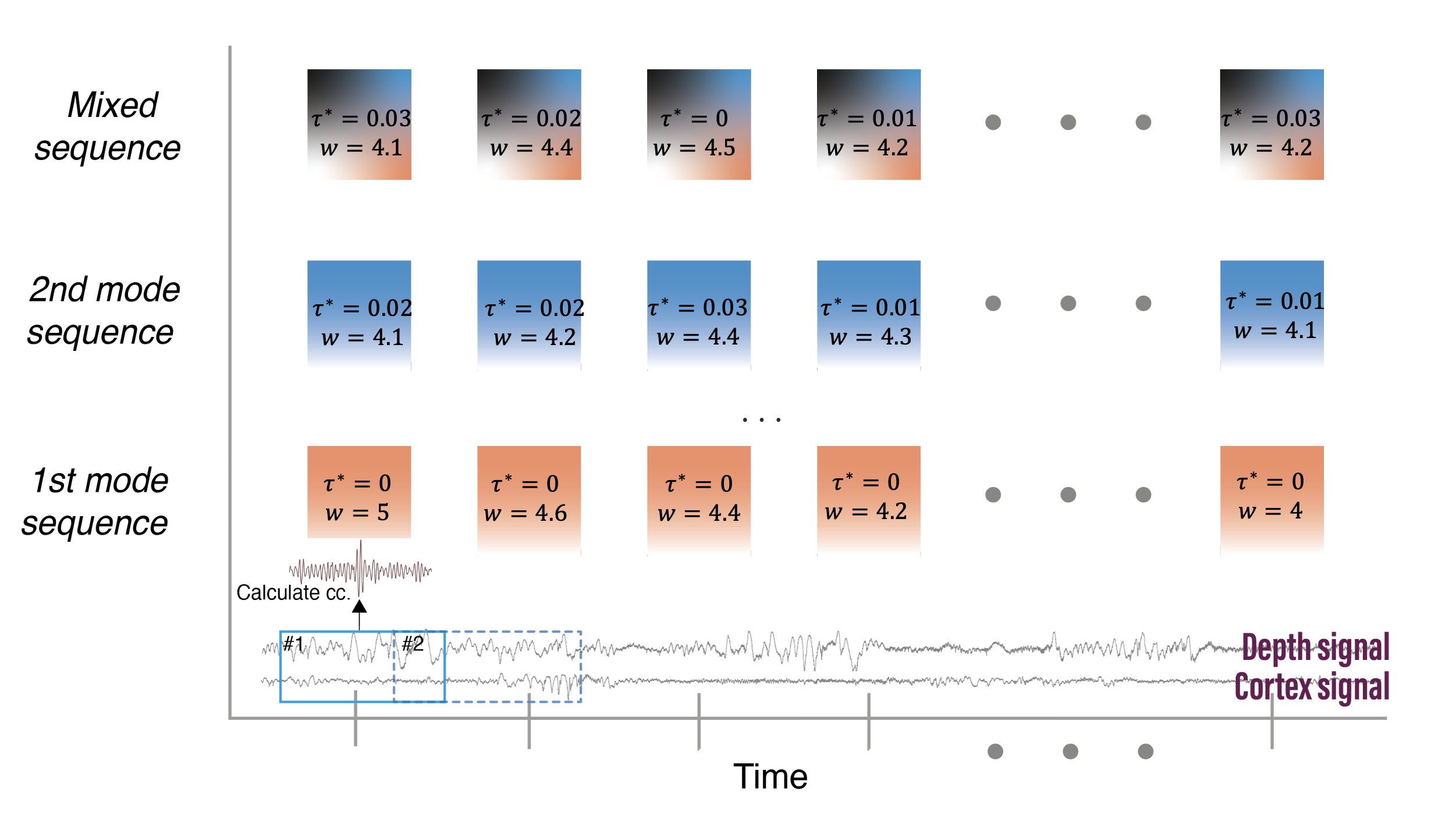}
	\caption{
		{\bf Illustration of the first, the second and the mixed modes sequences composition.} 
		This figure comprise of three rows and a bottom row which demonstrate both cortex and depth signals. The upper row cartoon describes a mixed sequence of different  $\tau^\ast$ realizations, the middle row describes a link consists of only second mode, and  the third row represents a link consisting of only the 1st mode sequence, all of length n. Note that in the third row  $\tau^\ast  = 0$ for all windows.	 
		A sequence is of duration $n$ time-windows in which each time window has $w>4.5$. That is, each sequence of length $n$, consists of $n$ time windows, each of which are the result of significant cross-correlated signal. 
		 We calculate the $\tau^\ast$ of each of the $n$ time windows, and ask what is the composition of those $\tau^\ast$'s, i.e, to what mode (one or two) this window belongs. 
		 }
	\label{fig:modes_illustration}
\end{figure}

 To study carefully the above question, we examined the links sequences compositions, i.e., the sub-links, by dividing the $\tau^\ast$ range in Fig.  \ref{fig:tauresults} into two regimes by the interception point between the two distributions, which we denote by ‘1’ if $\tau^\ast$ of the window belongs to the first low regime of $\tau^\ast$, and ‘2’ if it falls in the second high regime of $\tau^\ast$. 
Next, for each link sequence, we find the composition of its sub-links. For example, a sequence of 3 sub-links, may result in the composition ‘1,1,2’, meaning that the first and second windows had $\tau^\ast$’s from the first mode, and the third window had $\tau^\ast$ in the second mode regime, as illustrated in Fig. \ref{fig:modes_illustration}. Interestingly, we find that for most of the links, the whole sequence was composed by either consecutive ‘1’ or consecutive ‘2’. Meaning, that every interaction, i.e., link, is mainly of one type, and dominated by a single transfer mechanism. We assume that the first mode, which consists of $\tau^\ast\sim0$, are interactions that occur due to volume conduction process. This may fit the spread velocity of the bio-electromagnetic waves originating from the superposition of cortical neuronal activity and ion currents  \cite{nunez2006theoretical}.
The second mode of $\tau^\ast\sim0.02$ seconds probably represents information flow. 
 It can be seen in Fig.\ref{fig:modes_illustration} that only a very small fraction of the sequences are composed of mixed modes of 1’s and 2’s, (e.g., ‘1,2,1’). 

\FloatBarrier
\FloatBarrier

\begin{figure}[h]
	\centering
	\includegraphics[scale=0.6]{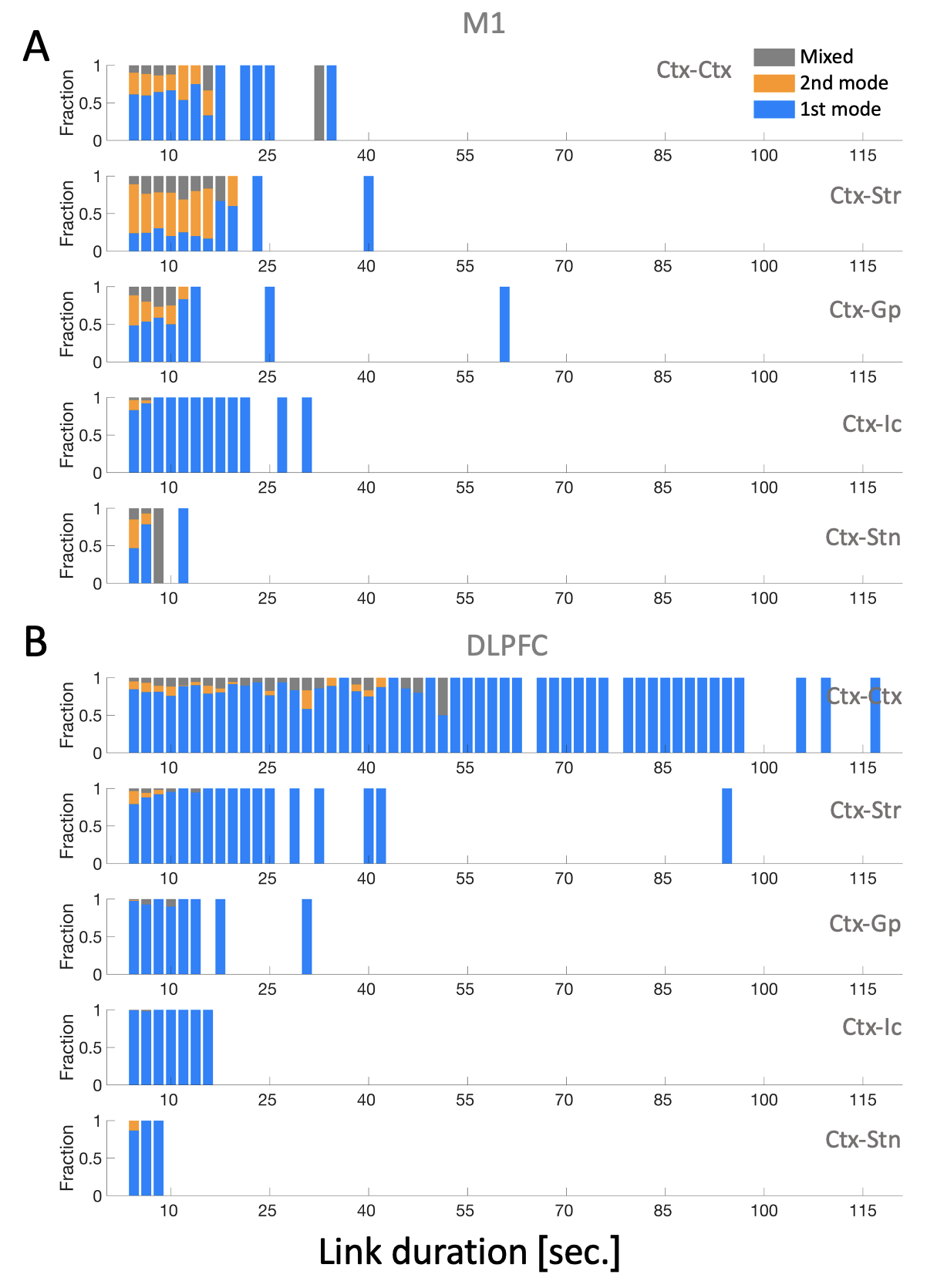}
	\caption{{\bf Sequences modes composition. } {\bf A,} Analysis of composition of links obtained between M1 electrodes and the deep electrode. {\bf B,} The composition of links between the DLPFC electrodes and the deep electrode. For each link of time length L, we analyze its 2.5 sec. sub-links $\tau\ast$ values, and test whether they are in the ranges of the first or second modes, seen in Fig. \ref{fig:tauresults}. We find, that for most of the sequences, the sequence belongs to the the 1st or 2nd modes exclusively, while only a very small fraction of the sequences contains a mix of the both modes. If the whole sequence contains only $\tau\ast$ values from the first (second) mode, we call it a "first (second) mode" sequence. If the sequence contains combinations of first and second modes, we say it is a "mixed" sequence. Note that the frontal cortex links are dominated by first mode sequences with $\langle\tau\rangle \backsimeq  0$, while second mode sequences $\langle\tau\rangle \backsimeq  0.02$ appear much more in the M1 links. Also note, that for the internal capsule (Ic) which consists of white matter, most links are from first mode in both frontal cortex and M1. 
		}
	\label{fig:modes_composition}
\end{figure}
\FloatBarrier

\section{Summary and discussion}
We found new local field potential (LFP) synchronization properties of links between the cerebral cortex (M1 and DLPFC) and the BG ( striatum, globus pallidus internal capsule and subthalamic nucleus). These features may represent a key to solving the enigma whether, BG LFP’s phenomena represent volume conduction or synaptic input (information flow) phenomena. We develop a novel cross-correlation method to identify significant links and define a duration of link pattern to address this question. We analyzed LFP, in the range of [0.1,80] Hz by seven micro-electrodes fixed on the deep layers in two regions of the  cerebral cortex: the primary motor cortex (M1) and the dorsolateral prefrontal cortex (DLPFC). One micro- electrode was recorded simultaneously in the basal ganglia from the DLPFC to the striatum, Globus pallidus, internal capsule and subthalamic nucleus. All recording done in a primate during spontaneous behavior in alert state.

Our results suggest  that the LFP signal combines two types of mechanisms, volume conduction and synaptic inputs  \cite{lalla2017local} .Each recording of 120 seconds (between the CTX and BG sites) has been divided into 2.5 seconds windows with overlapping of 0.25\% of the  window size. When analyzing the cross-correlations between the signals in both locations (CTX and BG), we found that the vast majority of windows were not correlated (defined by link-strength $w < 4.5$). As shown in Table 2, only 8\% to 14\% from the M1-BG pairs and 4\% to 51\% from the DLPFC-BG pairs can be regarded as significant links, or real-links. Additionally, we addressed the cross-correlation time-delay  which represents the time difference between the appearance of the signal in the two sites. The distributions of the significant links time delay (see figure 4) have been found to be bimodal; the first mode shows a peak  at $\tau^\ast \in 0 \pm0.015$ at both the M1-BG and the DLPFC-BG  pairs and the second mode at $\tau^\ast \in 0.03 \pm0.015$ mostly at the M1-BG pairs.

\subsection{ Differences in LFP’ links between each M1, DLPFC and BG }
We found differences in the cross-correlation links-properties between the M1-BG and DLPFC-BG, where M1-BG has a much more functional-links (information transfer links) of non-zero-time delays rather than between DLPFC-BG \cite{haber2003primate}. This may support the claim that the basal ganglia participate in functionally segregated circuits with motor and non-motor areas of the cerebral cortex, with a small bias to motor functionality
\cite{bostan2018functional}.

When we compare link time duration (or link length) from M1 and from the DLPFC, we observe that as the recorded nuclei is deeper, links time duration becomes shorter. Furthermore, links  between the DLPFC and BG are longer than their analogous between the M1 and BG, see Fig. \ref{fig:linklengths}. This loss of long duration links may be attributed to the longer distances which could introduce more noise in the signals.  
Furthermore, when looking at the $\tau^\ast$ distributions, we observed two main modes, see Fig. \ref{fig:tauresults}. One, characterized by a short time delays ($\tau^\ast \sim 0$), and the second, with longer time delays ($\tau^\ast \sim 0.02$).
We hypothesize, that these two modes are related to two different mechanisms of interaction, volume conductance (short $\tau^\ast$, 1st mode) and subthreshold phenomena as synaptic inputs - an information transfer (longer $\tau^\ast$, 2nd mode). 

Another important question we ask is whether a long term interaction, e.g. link, is mainly characterized by a $\tau^\ast$ sequence that belongs to one of the two modes. Indeed, we find that most of the links consist of mainly by either 1st or 2nd mode, and very little are a mix of the two, see Fig. \ref{fig:modes_composition}. If this is the case, it indicate further the existence of two mechanisms of transfer. Furthermore, the
Frontal cortex - deep electrode links comprise mostly of 1st mode sequences, while 2nd mode sequences are more prevalent in the M1 - BG, deep electrode links. 	

Logothesis et. al. \cite{logothetis2007vivo} performed in vivo measurement of cortical (visual cortex) impedance spectrum in monkeys: Impedance is a measure of the ability of a circuit to oppose the flow of electrical current. In biological tissues a “circuit” is the volume conductor, consisting of various intra- and extracellular components of the brain tissue, and the “current'' is generated by the movement of ions obeying Maxwell-laws. They showed that cortical impedance was independent of frequency, and was homogeneous and tangentially isotropic within the gray matter, which can be theoretically predicted assuming a behavior of a pure-resistive conductor. Logothesis et. al. \cite{logothetis2007vivo}  assumption may correspond to our analyses and results where we assume that the zero-lags cross correlated LFP’ segments may relate to volume conduction phenomenon that obey the Maxwell equations, while non-zero lags can be excluded from being volume conduction signals.

Our method and results may provide a framework to distinguish between volume conduction interactions, and real information flow, which currently is an unsolved question (XX-REFERNCES). Future work may use tools from information theory and test whether links that belong to the second mode, convey more information compared to those of volume conductance. Finally, as seen from our study, local field potentials are signals that are very easily obtained and studied. The current developed framework and study could open a wide range of future possibilities for deep brain stimulation (DBS) research and applications, for example in studying closed-loop implementation of BG-Cortex LFP based \cite{swann2018adaptive} and adaptive deep brain stimulation protocols.

 \newcommand{\newblock}{}
 
\bibliographystyle{unsrtnat}
\bibliography{lpf_bib.bib}
\clearpage

\section*{Supporting information}

\begin{figure}[h]
	\centering
	\includegraphics[scale=0.35]{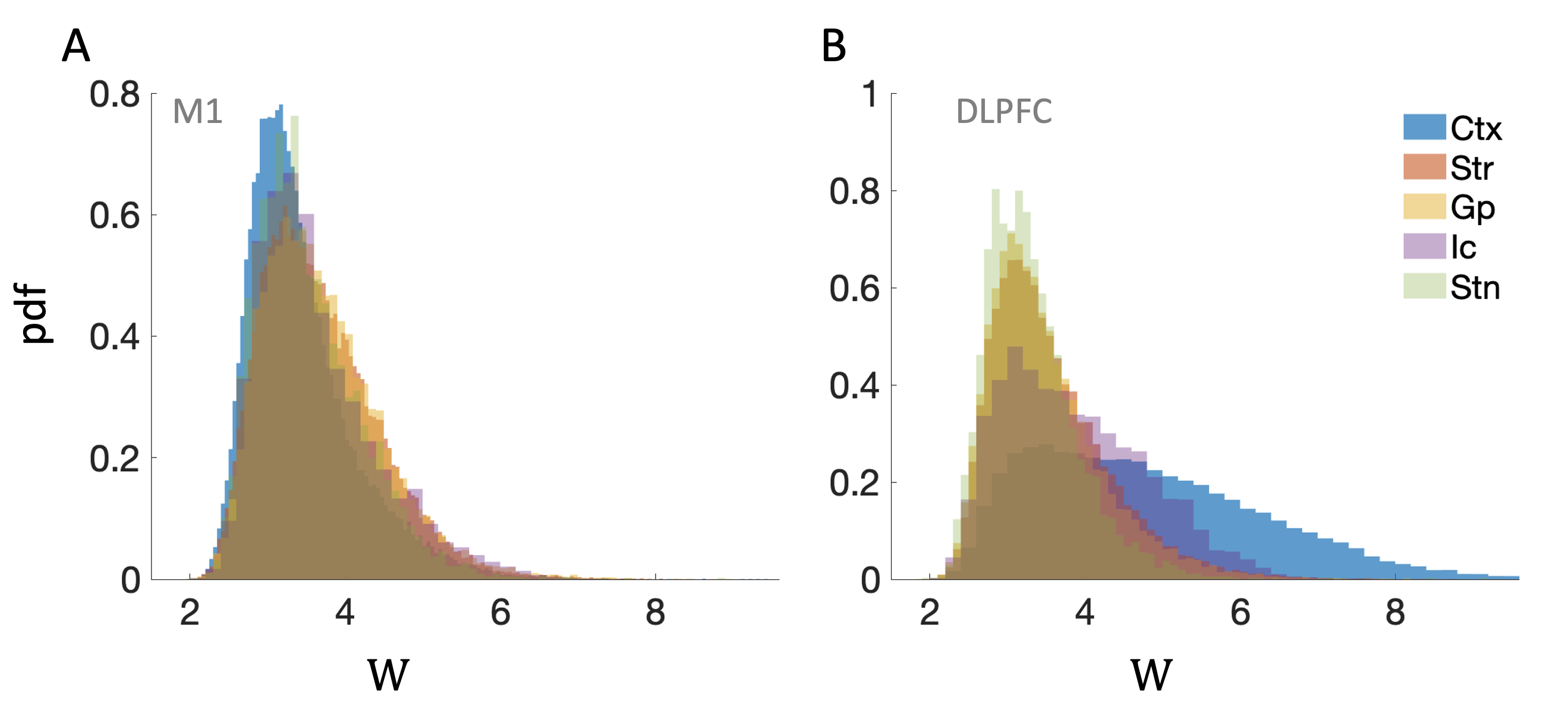} 
	\caption{ {\bf  $w$ distribution presented for M1 and DLPFC, per nuclei.}
			{ \bf A,} weights distribution of M1-Ctx, for the five nuclei. {\bf B,} same as A for the DLPFC-Ctx interactions.}
	\label{fig:SI_w_per_nuclei_dist}
\end{figure}

\begin{figure}[h]
	\centering
	\includegraphics[height=5.4cm ,width=8cm]{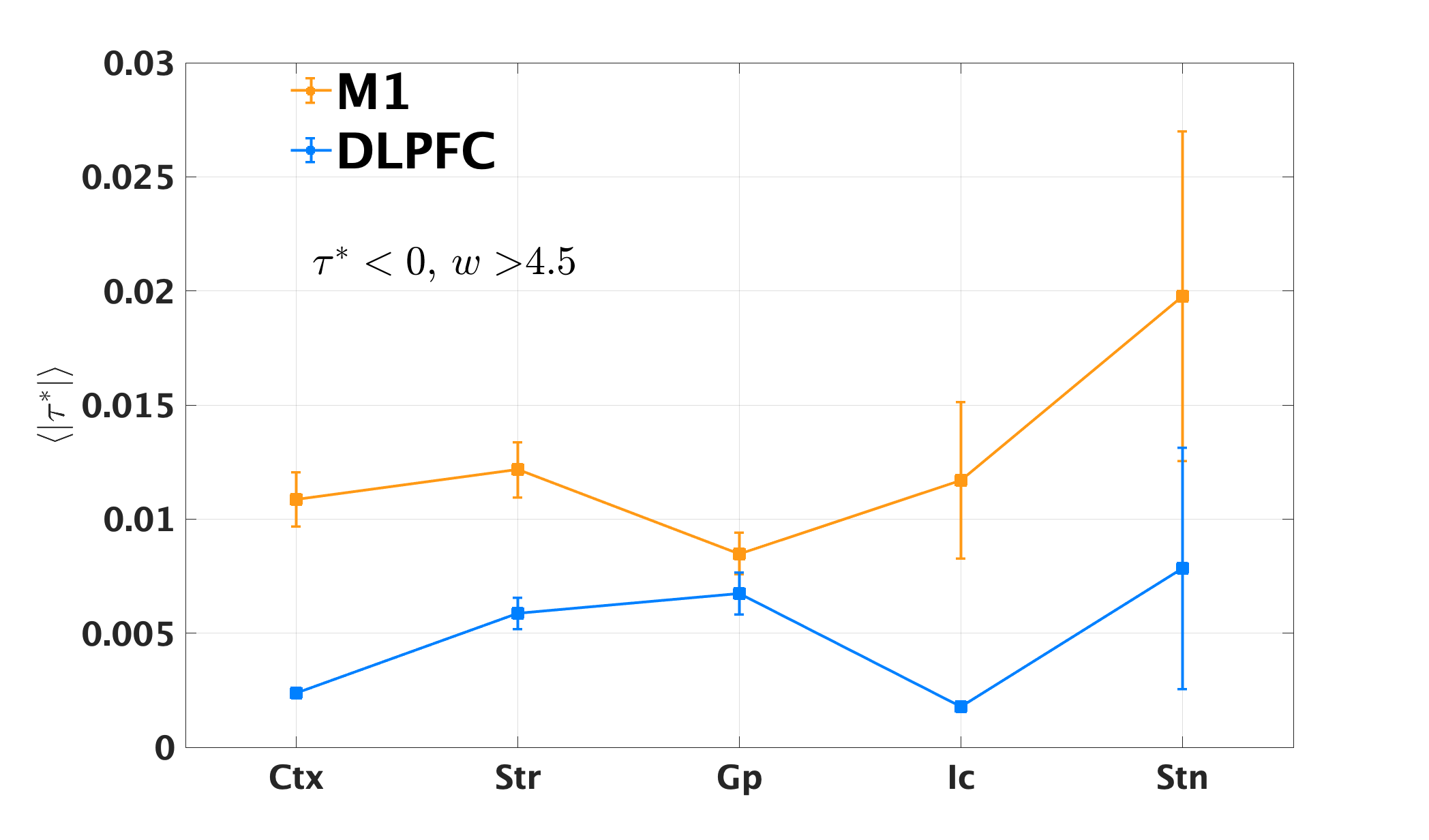} 
	\caption{{\bf Average time delay per nuclei, for interactions that were initiated in the basal ganglia. } \textbf{(a)} Average $|\tau^*|$'s of nuclei, of interactions in which $\tau<0$. We show the time delays of the two electrodes sets, M1 and DLPFC. $\tau^* < 0$ corresponds to the case in which the interaction initiated in the basal ganglia. Similar to the results that are shown in Fig.\ref{fig:tauresults}(A) M1 electrodes have a larger time delays, than the DLPFC with a similar profile. This holds for all nuclei. }
	\label{fig:SI_neg_t_per_nuclei_mean}
\end{figure}

\end{document}